\documentclass[12pt]{article}
\usepackage{amssymb,amsmath,amscd}
\usepackage{psfig}
\def\Rpar{{\mathbb R}_{\parallel}}
\def\Rper{{\mathbb R}_{\perp}}
\newcommand{\cF}{{\cal F}}
\newcommand{\cR}{{\cal R}}
\newcommand{\mb}[1]{{\mathbb {#1}}}

\begin{document}
\begin{center}
{\Large \bf
Noncommutative Torus from Fibonacci Chains \\ via Foliation \\
}
\vspace{12mm}
Hyeong-Chai Jeong${}^{a,}$\footnote{hcj@kunja.sejong.ac.kr}, \
Eunsang Kim${}^{b,}$\footnote{eskim@wavelet.hanyang.ac.kr}, \ and \;
Chang-Yeong Lee${}^{a, c,}$\footnote{leecy@kunja.sejong.ac.kr} \

\vspace{5mm}
${}^a${\it
Department of Physics, Sejong University, Seoul 143-747, Korea}\\
${}^b${\it Department of Industrial and Applied Mathematics,
Kyungpook \\ National University, Taegu 702-701, Korea}\\
${}^c${\it Theory Group, Department of Physics, University of
Texas, \\ Austin, TX 78712, USA} \\
\vspace{12mm}

\end{center}

\begin{center}
 {\bf Abstract}
\vspace{5mm}

\parbox{125mm}{
We classify the Fibonacci chains (F-chains) by their index sequences
and construct an approximately finite dimensional (AF) $C^*$-algebra
on the space of F-chains as Connes did on the space of Penrose
tiling. The K-theory on this AF-algebra suggests a connection between
the noncommutative torus and the space of F-chains. A noncommutative
torus, which can be regarded as the $C^*$-algebra of a foliation on
the torus, is explicitly embedded into the AF-algebra on the space of
F-chains. As a counterpart of that, we obtain a relation between the
space of F-chains and the leaf space of Kronecker foliation on the
torus using the cut-procedure of constructing F-chains.
}
\vfill
\end{center}
\setcounter{footnote}{0}

\pagebreak

\section{Introduction}

Recently, noncommutative geometry (NCG) has been one of the
most active areas in mathematics with increasing interest and
application to physics~\cite{landi97,Connes94B,Madore95B}.
Not only it opens new areas in pure mathematics
but its application to physics has now reached to the very frontier of
fundamental physics, such as string and M theories~\cite{Connes98,Seiberg99}.
Many in the  string/gravity  circle, now consider
NCG as a very possible candidate for the underlying mathematical
framework of quantum theory of gravity~\cite{Douglas99,Banks97,Madore98}.
However, applications of NCG to physical systems have not been
confined to high energy physics.
Bellissard already applied NCG to the quantum
Hall effect and explained the Hall conductivity using the K-theory on
the noncommutative algebra of functions on the Brillouin
zone~\cite{Bellissard86}.

Quasicrystals seem to be another novel systems to which we may able to
provide a ``quantum-jump'' progress when we adapt a NCG approach.
Quasicrystals are new types of solids with ordered atomic arrangement
but with a discrete point-group symmetry forbidden for periodic
systems. Discovery of such materials in 1984~\cite{Shechtman84} has
brought tremendous impact on condensed matter physics and
material science. By then, the only known ordered solid state
structures were crystals and solid states were considered either
periodically ordered crystals or disordered amorphous materials. The
structure of this new type of solids
has been explained with Penrose tiling in which two types of prototiles
arranged aperiodically~\cite{Penrose74}.
One can show that the Penrose tiling lattice has the translational
order and the rotational symmetry
of observed quasicrystals by calculating their Fourier
components~\cite{Levine84}.
Furthermore, Penrose tiling models provides clues to solving
the puzzle of physical realization of such structure through atomic
interaction, that is, the question of ``why do atoms form complex
Penrose tiling pattern rather than regularly repeating crystal
arrangement?''~\cite{Jeong97,Steinhardt96}.
However, the study of its dynamical properties, the most important
secret of quasicrystals, is in its infant stage yet. This may require
a new tool to analyze since quasicrystals defy
the standard classical classification of
solids. We think that NCG can be a candidate for this.
Connes already pointed out that the space of Penrose
tiling can be analyzed nontrivially only with noncommutative
algebra which has a quantum mechanical nature~\cite{Connes94B}.
An indication of quantum nature may have already appeared
in the fact that the symmetry of
Penrose tiling is not intuitively observed from its real space lattice
structure. As stressed by Rabson and Mermin~\cite{Rabson91}, the
symmetry of Penrose tiling is easily seen in Fourier transformed space
through the phases of the wave-functions in a scattering process, which
hardly play any role in the classical treatise.

Connes' analysis of the Penrose tiling space is based on the scale
invariance of the Penrose tilings. Using the inflation~(see
section~2) a Penrose tiling can be identified with a sequence
consisted of 0's and 1's~\cite{Connes94B,Grunbaum89B}.
Two different
sequences correspond to the same tiling if their
entries differ only in a finite number of terms.
When this equivalence relation is taken into account,
the space of tilings is given by the quotient space obtained from the
space of sequences mod out by the equivalence relation.
As Connes pointed out in his book~\cite{Connes94B},
one can hardly get any interesting information about this
space if it is treated as an ordinary space with classical tools.
For given any two Penrose tilings, one cannot distinguish one
from the other with any finite portion of them since it
appears in both tilings~\cite{Grunbaum89B}.
This tells us that the topology of the space of tilings is trivial,
namely the space of tilings is equivalent to a single point.
However, treating the space of tilings as a quantum space or
noncommutative space, one can find its interesting topological
invariant, the dimension group which is not trivial
at all~\cite{Connes94B}.
This is because a topologically trivial space cannot be described
nontrivially by complex-valued functions. However, with
operator-valued functions on this space one can explore the nontrivial
structure of this seemingly trivial space.

The study of quasiperiodic structure along the noncommutative geometric
approach was first done by Bellissard {\it et al.}~\cite{Bellissard90p}
in a one-dimensional (1D) case. They investigated its spectral
properties and tried to construct a quantal observable algebra
which plays the role of the above mentioned
operator-valued functions.
However, their investigations fell short of geometric
properties in the sense of Connes.

Recently, the study in the noncommutative geometry framework
was done by Landi and companies~\cite{Ercolessi98} from the
view point of noncommutative lattice which can be regarded
as a finite topological approximation of a quantum physical model.
They performed their investigation by studying the K-theory
of the approximately finite dimensional
(AF) $C^*$-algebra. For the Penrose tiling case, they retrieved the
Connes' result.

However, so far not much has been known about the underlying nature
of the space of tilings. On the other hand, one can see a
close resemblance of the K-theory result of Connes to that of the
noncommutative torus.

In this paper, we investigate this aspect of the tiling space.
We analyze the space of Fibonacci chains (F-chains) which is
isomorphic to the space of Penrose tilings (see section~2).
Using the ``projection'' method explained in section~3,
both Fibonacci chains and
Penrose tilings can be represented as points in the higher dimensional
torus. However, we choose the space of F-chains (rather than
the Penrose tiling space) since its geometrical interpretation is
simpler in the torus representation~\cite{Baake97}. By analyzing this
torus parameterization from the perspective of foliation,
which has become an important tool for the investigation of
noncommutative geometric property, we show that the parameterized
torus can be foliated to make a noncommutative torus and
explain why
the map from the F-chains to the leaves of foliation
is surjective.

The organization of the paper is as follows.
In section~2, we introduce the deflation method of obtaining the
F-chains. We then construct the index sequences of
the F-chains and explain the equivalence relation on them.
With this, we construct an AF-algebraic structure in the manner that
Connes formulated on the space of the Penrose tilings and calculate
the K-theory in this scheme.
In section~3, we ``lift'' the F-chains to a two-dimensional (2D)
hyperspace. This procedure naturally leads to the torus
parameterization of the F-chains. We then show that the torus
parameterization becomes the Kronecker foliation on the 2-torus when the
equivalence relation of F-chains is applied.
This mapping from the space of F-chains to the leaf space of
the foliation is surjective. There is one ``singular'' leaf which
corresponds to two different classes of F-chains.
This ``singularity'' is explained in terms of both the projection method
and the cut-procedure of obtaining the F-chains.
In section~4, we extend the leaf space such that it can be
isomorphic to the space of F-chains and embed the $C^*$-algebra
of leaves of foliation into an AF-algebra on this extended space.
We first obtain the equivalence relations on the extended leaf space
using the equivalence relation of corresponding
F-chains~\cite{Connes94B} in the finite steps.
This equivalence relation partitions the space to the
finite intervals. The AF-algebra is obtained as an
inductive limit of the finite algebra on the space
of the finite intervals. In our concluding remarks in
section~5, we summarize our results and discuss the implication for
future research in the properties of Penrose tiling.

\vspace{1mm}
\section{Fibonacci chain and its K-theory}
\label{Fibo_chain}
A typical example of a one-dimensional (1D) quasiperiodic structure
is the so called Fibonacci chain (F-chain). An F-chain is a special
infinite sequence of two segments, say, one short
$S$-segment and one long $L$-segment with following properties;
\begin{enumerate}
\item Any finite part of the sequence appears infinite times but none
      of them are consequently repeated more than two times.
\item One type of segment (say $S$) cannot consequently repeated
      ($SS$ is not allowed).
\end{enumerate}
One way to obtain this sequence of segments is the ``deflation''
method~\cite{Grunbaum89B}. In this method, we start from a finite
sub-chain of an F-chain. We then operate iterative substitution
(deflation) rule, $S\rightarrow L$ and $L \rightarrow LS$ to build
successive strings with increasing length. At any point in the chain,
the type of segment ($L$ or $S$)
is uniquely determined by the chosen starting sequence. Figure 1
shows such successive iterations when the starting sequence is just
one segment $L$. An infinite number of iterative
deflations produce an F-chain.

\vspace{1mm}
\begin{figure}[!t]
\centerline{\psfig{file=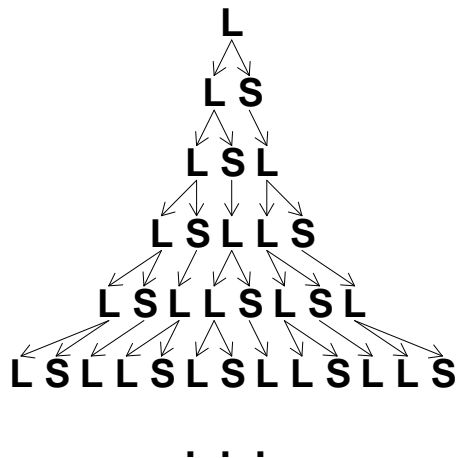,width=70mm}}
\caption{A way of constructing F-chain
using deflation. In each deflation step, every
$S$ segment is replaced by $L$ and
every $L$ segment is replaced by $LS$.
An infinite number of iterative
deflations produce an F-chain.
}
\label{f-1}
\end{figure}

\begin{figure}[!t]
\centerline{\psfig{file=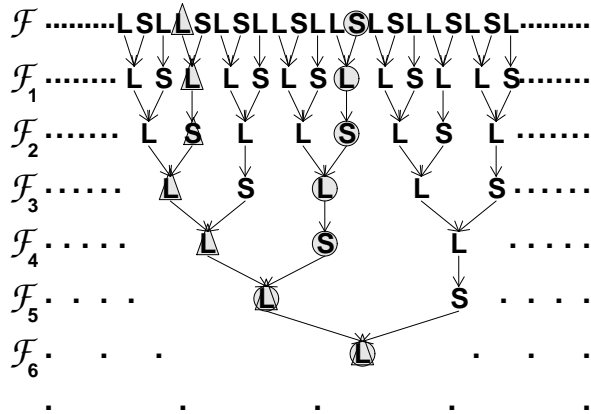,width=90mm}}
\caption{Successive inflations of the F-chain ${\cal F}$
produce a sequence of F-chains ${\cal F}_1$,
${\cal F}_2$, and so on.
For the segment denoted by the triangle in ${\cal F}$,
the index sequence is given by $(0,0,1,0,0,0,\ldots)$
whereas that for the segment denoted by the circle
is given by $(1,0,1,0,1,0,0,\ldots)$.
}
\label{f-inf}
\end{figure}

The inverse process of the ``deflation'' is the ``inflation''.
Now we begin with a F-chain $\cF$ of two segments
$L$ and $S$ and apply a composition (inflation), $LS \rightarrow L$
and $L\rightarrow S$. This produces another F-chain
${\cal F}_1$ of the two segments of $L_1$ and $S_1$, where
$L_1 = LS$ and  $S_1 = L$.
Successive application of the compositions
yields a series of F-chains
${\cal F}_n$ of two segments $L_{n}$
and $S_{n}$ where $L_{n} =  L_{n-1} S_{n-1}$
and $S_{n} =  L_{n-1}$ with
$L_0 := L$ and $S_0 := S$
as shown in Fig.~\ref{f-inf}.

This naturally introduces the index sequences
of the chains~\cite{Grunbaum89B}.
For a given segment $\alpha$ in the original
F-chain ${\cal F}$,
the index sequence $i({\cal F},\alpha)$ is defined as
an infinite sequence of integers $(a_0,a_1,a_2,\ldots)$
where $a_n = 1$ or $0$ according as whether $\alpha$ belongs to an
$S_n$ or $L_n$ segment in ${\cal F}_n$ for $n=0,1,\ldots$.
~From the inflation rule
($LS \rightarrow L$, $L\rightarrow S$),
it is clear that
an $S$ segment in ${\cal F}_n$ must belong to
an $L$ segment in ${\cal F}_{n+1}$, that is,
$a_n\!=\!1$ implies $a_{n+1}\!=\!0$ for an
index sequence $(a_n)$.
In fact, one can show that the set of
index sequences of F-chain $i({\cal F},\alpha)$
is isomorphic to the set $Z$ of sequences $(a_n)$,
with $a_n = 1$ or 0 such that
$a_n\!=\!1\ \Longrightarrow \ a_{n+1}\!=\!0$~\cite{Grunbaum89B,expl_cantor}.

Figure~\ref{f-inf} illustrates the way of constructing
the index sequence using inflation.
For the segment denoted by the triangle in ${\cal F}$,
the index sequence $(a_n)$ is given by
$(a_n) = (0,0,1,0,0,0,\ldots)$
since this  segment belongs to $L,L,S,L,L,L$ segments in
the ${\cal F}$, ${\cal F}_1$, ${\cal F}_2$, ${\cal F}_3$,
${\cal F}_4$ and ${\cal F}_5$ chains respectively.
Similarly, the index sequence $(b_n)$ for the segment denoted
by the circle is given by $(b_n) = (1,0,1,0,1,0,0,\ldots)$.
Note that the indices in both sequences are the same for $n\ge 5$ since
both the triangle and circle segments in ${\cal F}$ belongs
to the same segments for $n\ge 5$ chains.
In fact, the inflation will make any two segments in ${\cal F}$
separated by a finite distance  belong to the same segment in
${\cal F}_n$ for sufficiently large $n$.
Therefore, for the index sequences,
$(a_n) = i ({\cal F},\alpha)$
and
$(a'_n) = i ({\cal F},\alpha')$
from two given segments in the same chain,
there must be an integer $M$ such that
$a'_n = a_n$ for all $n > M$.
This naturally leads to the following definition of
the equivalence relation $\cR$ on $Z$;
\begin{eqnarray}
(a_n) \sim (a_n')
      &\text{iff}&\text{there is an integer } M>0 \nonumber \\
      &          &\text{such that}\ a_n=a_n' \text{ for all } n>M.
\label{e-equivR}
\end{eqnarray}
With this equivalence relation, it is trivial that
any two index sequences from the same F-chain
are in the same equivalence class.

Conversely, one can also show that any two different sequences in $Z$ with
$a_n=a_n'$ for all $n>M$ can be constructed as index sequences from two
different segments in the same F-chain~\cite{Grunbaum89B}.
Therefore, the space of F-chain is given
by the quotient space $X=Z/\cR$. In fact this identification
allows us to see the space of F-chains as a noncommutative space, as
was noted in~\cite{Connes94B}. In other words one can define the
$C^*$-algebra associated to the quotient space $X$.

In what follows we review the construction of the $C^*$-algebra
and the computation of its $K$-theory
following the lines of ~\cite{Connes94B}.
Consider the set
$$Z_n=\{(a_0,\cdots,a_n)\mid a_j\in\{0,1\}\text{ and }
a_j=1\Longrightarrow a_{j+1}=0\}.$$
These sets form an inverse system of sets:
$$\cdots \longrightarrow Z_{n+1}\longrightarrow
  Z_n\longrightarrow \cdots \longrightarrow Z_1$$
under the projection maps
$Z_{n+1}\longrightarrow Z_n$ given by
$(a_0,\cdots,a_n,a_{n+1})\mapsto (a_0,\cdots,a_n)$.
Note that
the inverse limit $\varprojlim Z_n=Z$ is simply
the set of  all F-chains.
On each $Z_n$, there is  an equivalence relation $\cR_n$ given by
\begin{eqnarray}
 (a_0,\cdots,a_n) \sim (a_0',\cdots,a_n')
                  &\text{ \ iff \ }&    a_n=a_n'.
\label{e-equivRn}
\end{eqnarray}
Let $X_n=Z_n/\cR_n$ be the set of all equivalence classes.
Since each entries of sequences in $Z_n$ are either $0$ or $1$,
there are only two elements in $X_n$.
Those two elements correspond to $0$ or $1$
in the final entry. Thus the space $X_n$ cannot be described
non-trivially by means of functions with values in complex
numbers, $\mathbb C$.  Instead, if we take operator-valued functions on $X_n$,
there exists a very rich class of such functions. For this, each $[x]\in
X_n$, one can associate a finite dimensional Hilbert space $l^2_x$
having elements of $[x]$ for an orthonormal basis and
the algebra is given by the set of all
functions on $X_n$ with values in operators on $l^2_x$.
Note that if the dimension of $l^2_x$ is $k$, then
the algebra of operators on $l^2_x$ is the algebra of all $k\times k$
matrices.
More
explicitly, if $x_0$( $x_1$ resp.) represents the class in $Z_n$
with $0$ ($1$ resp.) in the final entry, then the dimension of $l^2_{x_0}$
($l^2_{x_1}$ resp.) is  is the number of distinct elements in $Z_n$ that end with $0$
($1$ resp.). Let $k_n$ and $k'_n$ be the dimension of $l^2_{x_0}$
and $l^2_{x_1}$, respectively.
Now the algebra of functions on $[x_0]$ ($[x_1]$
resp.) with values in $M_{k_n}(\mathbb C)$
($M_{k_n'}(\mathbb C)$
resp.) is simply
$M_{k_n}(\mathbb C)$ ($M_{k_n'}(\mathbb C)$ resp.) and thus
the $C^*$-algebra $A_n$ of operator-valued functions on $X_n$ is
identified with $M_{k_n}(\mathbb C)\oplus M_{k_n'}(\mathbb C)$.
Also we have an
inclusion map
$A_n\longrightarrow A_{n+1}$ and it is uniquely determined by the
equalities
\begin{eqnarray}
\begin{pmatrix} k_{n+1}\\k_{n+1}'\end{pmatrix}=\begin{pmatrix}
1&1\\1&0\end{pmatrix}\begin{pmatrix} k_n\\k_n'\end{pmatrix}.\label{inc-eq}
\end{eqnarray}
It allows embedding $A_n$ as block matrices in $A_{n+1}$ i.e.,
$$\begin{pmatrix}
M&0\\
0&N\end{pmatrix} \mapsto\begin{pmatrix} M&0&0\\0&N&0\\0&0&M\end{pmatrix}$$
where $M\in M_{k_n}(\mathbb C)$ and $N\in M_{k_n'}(\mathbb C)$.
Now we have an inductive system of $C^*$-algebras:
\begin{eqnarray}
A_1\longrightarrow A_2\longrightarrow \cdots \longrightarrow A_n
\longrightarrow \cdots
\label{ind-sys}
\end{eqnarray}
Let $A=\varinjlim A_n$ be the inductive limit of the system.
Then
 $A$  is an AF-algebra and is considered as the $C^*$-algebra of
 $X$.
In general, approximately finite $C^*$-algebra or
AF-algebra is defined by an inductive limit of a
sequence of finite-dimensional $C^*$-algebras and such algebras
can be completely classified by its $K$-theory~\cite{Blackadar86B}.
By applying  basic properties of
K-theory~\cite{landi97,Blackadar86B,Wegge93B} to the system~(\ref{ind-sys}),
one can see that the $K$-theory of $A$ is also
determined by the equations~(\ref{inc-eq}). Note that for each
$n$,
$$K_i(A_n)=K_i(M_{k_n}(\mathbb C)\oplus M_{k_n'}(\mathbb
C))=\bigl\{\aligned \mathbb Z\oplus\mathbb Z \text{ \ \ \ if
}i=0\\ 0\text{ \ \ \ \ \ if }i=1,\endaligned
$$
and the positive cone is given by
$$K_0^+(A_n)=\mathbb Z^+\oplus \mathbb Z^+.$$
The map $K_0(A_n)\to K_0(A_{n+1})$ is uniquely determined by
the equation~(\ref{inc-eq}) and is represented by
$\begin{pmatrix}1&1\\ 1&0\end{pmatrix}$.
Since $\begin{pmatrix} 1&1\\1&0\end{pmatrix}$ is invertible in
${\mathbb Z}\oplus{\mathbb Z}$, it is an isomorphism on
$K_0(A_n) \longrightarrow K_0(A_{n+1})$ 
for all $n\ge 0$ and we have
$$K_0(A)=\varinjlim K_0(A_n)\cong {\mathbb Z}\oplus{\mathbb Z}.$$
On the other hand, $\begin{pmatrix} 1&1\\1&0\end{pmatrix}$ is not
invertible in ${\mathbb Z}^+\oplus{\mathbb Z}^+$. To compute
$K_0^+(A)$, let $(a,b)~\in~{\mathbb Z}\oplus{\mathbb Z}$ and
then
$$K_0^+(A_1)=\{(a,b)\in {\mathbb Z}\oplus{\mathbb Z}
\mid a+b\ge 0 \text{ \ and  \ } b\ge 0\},$$
and if we let $\begin{pmatrix} 1&1\\1&0\end{pmatrix}^n=
\begin{pmatrix} f_n^{11}&f_n^{12}\\f_n^{21}&f_n^{22}\end{pmatrix}$, then
$$K_0^+(A_n)=\{(a,b)\in{\mathbb Z}\oplus{\mathbb Z}\mid
f^{11}_na+f_n^{12}b\ge 0 \text{  \ and \ }f_n^{21}a\ge 0\}.$$
~From the computation of
$ \begin{pmatrix} 1&1\\1&0\end{pmatrix}^n
 =\begin{pmatrix} f_n^{11}&f_n^{12}\\f_n^{21}&f_n^{22}\end{pmatrix}$,
we see that
$$\begin{pmatrix} 1&1\\1&0\end{pmatrix}^n
 =\begin{pmatrix} f_{n+1}&f_n\\f_n&f_{n-1}\end{pmatrix}$$
with the defining relation:
$$f_{n+1}=f_n+f_{n-1}, \text{ \ \ and \ \ }  f_1 = f_2 = 1.$$
Thus
$$\aligned
K_0^+(A_n)&=\{(a,b)\in {\mathbb Z}\oplus{\mathbb Z}
\mid  f_na+f_{n+1}b\ge 0 \text{ \ \ and \ } f_{n-1}a+f_nb\ge 0\}\\
&=\{(a,b)\in {\mathbb Z}\oplus{\mathbb Z}
\mid a+\frac{f_{n+1}}{f_n}b\ge 0\}.\endaligned$$
~From this
$$K_0^+(A)=\varinjlim K_0^+(A_n)=\{(a,b)\in {\mathbb Z}\oplus{\mathbb Z}
\mid a+\tau b\ge 0\},$$
where $\lim_{n\to\infty}\frac{f_{n+1}}{f_n}=\tau$ is the Golden
mean.
Now the space of F-chains is completely characterized by
the ordered group
$$(K_0(A), K_0^+(A)) =
({\mathbb Z}^2,\{(a,b)\in {\mathbb Z}\oplus{\mathbb Z}\mid a+\tau b\ge
0\}).$$

Recall that the noncommutative torus is the $C^*$-algebra
generated by two operators $u$, $v$ subject only to
$$uv=e^{2\pi i \Theta}vu$$
where $\Theta$ is a real number.
It is well-known that the $K$-theory of the noncommutative 2-torus
$A_\Theta$ is given by
$K_i(A_\Theta)\cong{\mathbb Z}^2$,
where $i=0,1$. In particular, $K_0(A_\Theta)$ is
isomorphic to ${\mathbb Z}\oplus\Theta{\mathbb Z}$ as ordered
groups by a theorem of Pimsner and Voiculescu~\cite{Pinsner80}.
Furthermore, the noncommutative torus can be embedded into
a certain type of AF-algebra as discussed in  Landi, Lizzi, and
Szabo recently~\cite{landi99}. In the above
we have
shown that the $C^*$-algebra of the space of F-chains is
an AF-algebra and its
$K$-groups were computed. Furthermore, the Bratteli diagram of
the AF-algebra  satisfies the
condition required by Landi {\it et al.}'s work with $c_n=1$ in their
notation~\cite{Ercolessi98,landi97}.  Thus one might
expect that the torus $A_\tau$ can be embedded into the
$C^*$-algebra of the space of F-chains. As a dual picture,
if we can realize the
noncommutative torus as a geometric object, then we may
characterize the space of F-chains from the space associated to
the algebra $A_\tau$. In the next section we will show
that the space of F-chains can be determined by
submanifolds of ordinary torus $\mathbb T^2$.

\section{Torus representation and foliation}
\label{s-TR}

In this section we establish a precise relation between the space of
F-chains and the leaf space of the Kronecker foliation on the torus.
This will be done in the torus representation of F-chains~\cite{Baake97}.
We first review the definition of the Kronecker foliation
and study the correspondence between the noncommutative torus and the
Kronecker foliation. We then show how they are related to the
space of F-chains.

In general, a foliation of codimension $q$ on an $n$-dimensional
manifold is a partition of the manifold into $p$-dimensional connected
submanifolds, where $n=p+q$. Such submanifolds are called the leaves
of the foliation.
Locally the leaves look like a set of parallel planes of
codimension $q$ in Euclidean space. The space of leaves can be
understood as families of solutions of systems of differential
equations and the study of foliation is the study of the global
behavior of such solutions. For example, a first order
differential equation is a vector field. For a vector field
without zeros, the orbits of the flow generated by the vector
field form a 1D foliation. See~\cite{Lawson74} for details
for the theory of foliations.

It is well-known that the 2-torus $\mathbb T^2$ is the only
oriented compact 2-dimensional manifold which admits a
non-singular codimension 1 foliation. Up to topological
equivalence one can classify smooth foliations
of $\mathbb T^2$~\cite{Lawson74}. In particular, there is a foliation which
contains no closed leaves and this foliation is equivalent to the
Kronecker foliation with irrational slope. Let
${\mathbb T}^2=S^1\times S^1={\mathbb R}^2/{\mathbb Z}^2$
with natural coordinates $(x,y)\in {\mathbb R}^2$. For non-zero
constants $a_1$ and $a_2$, a smooth one-form
$\omega=a_1\,dx+a_2\,dy$
on the torus defines a foliation on
$\mathbb T^2$.
The leaves of this foliation are the solutions of the
differential equation
$$dy=-\frac{a_1}{a_2}\,dx.$$
If $\frac{a_1}{a_2}$ is rational, then each leaf is closed and hence a
circle on the torus. If $\frac{a_1}{a_2}$ is irrational, then all the
leaves are diffeomorphic to $\mathbb R$ and each leaf is dense in
$\mathbb T^2$. This foliation is called the Kronecker foliation associated
to a real number $-\frac{a_1}{a_2}$. From now on, we will restrict ourselves
to the case when $-\frac{a_1}{a_2}=\frac{1}{\tau}$, where $\tau$ is the
Golden mean.
Each leaf in this case can be seen as a straight line
in $\mathbb R^2$ with the fixed slope, $y=\frac{1}{\tau} x+b$.
Since a straight line $y=\frac{1}{\tau} x+b$
is determined by its $y$-cut, we see that
the space of leaves of the foliation is parameterized by
the $y$-cuts. On the torus, two lines $\frac{1}{\tau} x+b$ and
$\frac{1}{\tau} x+b'$
represent the same leaf if $b-b'=\frac{1}{\tau} n$,
for some integer $n$.
This defines an equivalence relation on the $y$-cuts
and the leaf space $X_{\cF}$ of the Kronecker
foliation can be identified with the set of equivalence classes.
The topology on the space of leaves is the same as the quotient
topology of $S^1=\mathbb R/\mathbb Z$ divided by the partition
into orbits of the rotation given by $z\mapsto z+\frac{1}{\tau}$,
where $z\in S^1$, and hence there are no open sets in
$X_{\cF}$ except $\emptyset$ and $X_{\cF}$.
Therefore, the leaf space has the trivial topology as in the case of
the space of F-chains.

The Kronecker foliation can also be obtained from the {\it suspension
of diffeomorphisms}~\cite{Lawson74}.
Let $\psi_\tau:S^1\rightarrow S^1$ be the
diffeomorphism which is the rotation through angle
$\frac{2\pi}{\tau}$,
i.e., $\psi_\tau(z)=e^{\frac{2\pi i}{\tau}}\cdot z$,
\ $z\in S^1$.
The product manifold $S^1\times \mathbb R$
is foliated by the leaves of the form $\{z\}\times \mathbb R$.
This foliation on $S^1\times \mathbb R$ is invariant under the $\mathbb
Z$-action on $S^1\times \mathbb R$: for $z\in S^1$, $b\in \mathbb
R$,
\begin{eqnarray}
(z,b)^n=(\psi_\tau^n(z),b+n), \text{ \ \ \ } n\in \mathbb Z.
\label{action1}
\end{eqnarray}
This means that the quotient
$S^1\times_{\mathbb Z}\mathbb R\cong \mathbb
T^2$ carries a 1D foliation whose leaves are the image
of $\{z\}\times \mathbb R$ under the quotient map
$S^1\times \mathbb R\to \mathbb
T^2$. Equivalently, the leaves are transverse to the fibers of
$S^1\times_{\mathbb Z}\mathbb R\to \mathbb R/\mathbb Z\cong S^1$.
Thus the space of leaves of the foliation on $\mathbb T^2$ are
parameterized by $\mathbb R$ together with the $\mathbb Z$-action
associated to the action of Eq.~(\ref{action1}). This is exactly the same
relation as the one in the Kronecker foliation and hence the
foliation obtained via the diffeomorphism $\psi_\tau$ is
equivalent to the Kronecker foliation on $\mathbb T^2$ with the
irrational slope $\frac{1}{\tau}$. This is in fact followed by the
Denjoy's theorem which asserts that if a foliation of ${\mathbb
T}^2$ does not have compact leaves, then it is  topologically equivalent
to a foliation obtained by a suspension of an irrational rotation of the
circle. Also, Denjoy constructed examples of foliations on ${\mathbb
T}^2$ with exceptional minimal sets  and this motivated the study
of minimal sets of foliations of codimension one on compact
manifolds of dimension $\ge 3$. Here we briefly review Denjoy's
example which is obtained by the suspending the diffeomorphism $\psi_\tau:S^1\to
S^1$ with an exceptional minimal set \cite{Camacho85}. In section~4,
we will identify the set of all F-chains with the exceptional
minimal set. A subset $E$ of $S^1$ is said to be $minimal$ if it is
closed, nonempty and invariant under $\psi_\tau$ and also if $E'\subset
E$ is closed, invariant subset then either $E'=\emptyset$ or
$E'=E$. A minimal set is called $exceptional$ if it is
homeomorphic to a subset of the Cantor set on $S^1$. The
exceptional minimal set for $\psi_\tau$ is constructed in the
following manner. First, cut the circle $S^1$ at all the points of
an orbit $\{\theta_n\mid n\in {\mathbb Z}\}$ of the given irrational
rotation. At the $n$ cutting point, insert a segment $J_n$ of length
$l_n$ with $\sum l_n <\infty$. Then we get a new circle  and the set
$S^1-\cup_{n\in{\mathbb Z}}J_n=E$ is homeomorphic to the Cantor
set and this is the desired  exceptional minimal set.

~From the above construction of the Kronecker foliation, one can
relate the $C^*$-algebra of the foliation to the noncommutative torus.
Let $C(S^1)$ be the
$C^*$-algebra of continuous functions on $S^1$. Then the rotation
$\psi_\tau:S^1\to S^1$ induces the automorphism $\psi_\tau^*:C(S^1)\to C(S^1)$
given by $\psi_\tau^*(f)=f\circ\psi_\tau$, where $f\in C(S^1)$, or
\begin{eqnarray}
 (\psi_\tau^*f)(z)=(f\circ\psi_\tau)(z)
  = f(e^{\frac{2\pi i}{\tau}}\cdot z),\text{ \ \ \ }z\in S^1.
\nonumber
\label{rotation}
\end{eqnarray}
Let us denote the group of all automorphisms of $C(S^1)$
by $\text{Aut}(C(S^1))$.
Then the action of Eq.~(\ref{action1}) can be given as the
group homomorphism $\alpha:\mathbb Z\to \text{Aut}(C(S^1))$ which
is given by
\begin{eqnarray}
(\alpha(n)f)(z):=(\alpha_nf)(z)=(f\circ\psi_\tau^n)(z)=f(e^\frac{2\pi
ni}{\tau}\cdot z).
\nonumber
\label{action2}
\end{eqnarray}
Now the $C^*$-algebra of this action is so called the {\it crossed
product} $C^*$-algebra
$C(S^1)\rtimes_\alpha\mathbb Z$~\cite{Blackadar86B}.
As we have seen above,
this algebra is generated by the rotation and the $\mathbb
Z$-action. More explicitly, the $C^*$-algebra is represented on $L^2(S^1)$
with generators $U$ and $V$ according to the rotation and $\mathbb
Z$-action:
\begin{eqnarray}
(Uf)(z)=zf(z) \text{ \ \ and \ \ } (Vf)(z)
       =f(e^\frac{2\pi i}{\tau}\cdot z),
       \text{ \ \ \ }f\in L^2(S^1), \ z\in S^1.
\nonumber
\end{eqnarray}
It is easy to verify that the operators $U$ and $V$ satisfy the
relation
$$UV=e^\frac{2\pi n i}{\tau} VU.$$
Hence the $C^*$-algebra $C(S^1)\rtimes_\alpha\mathbb Z$ is
identified with the noncommutative torus $A_{\frac{1}{\tau}}$ and
also it can be regarded as the $C^*$-algebra on
the leaf space of the Kronecker foliation with the slope
$\frac{1}{\tau}$. Also this is Morita equivalent to the
noncommutative torus $A_\tau$~\cite{Connes94B} and hence its $K$-theory is given by
$K_i(A_\tau)\cong{\mathbb Z}^2$,
where $i=0,1$. In particular, $K_0(A_\tau)$ is
isomorphic to ${\mathbb Z}\oplus\tau\mathbb Z$ as ordered
groups as discussed in section 2. In below we will establish a
relation between the space of the F-chains and the leaf space of the
Kronecker foliation on the torus ${\mathbb T}^2$ appearing in the
torus representation~\cite{Baake97}.

\begin{figure}[!t]
\centerline{\psfig{file=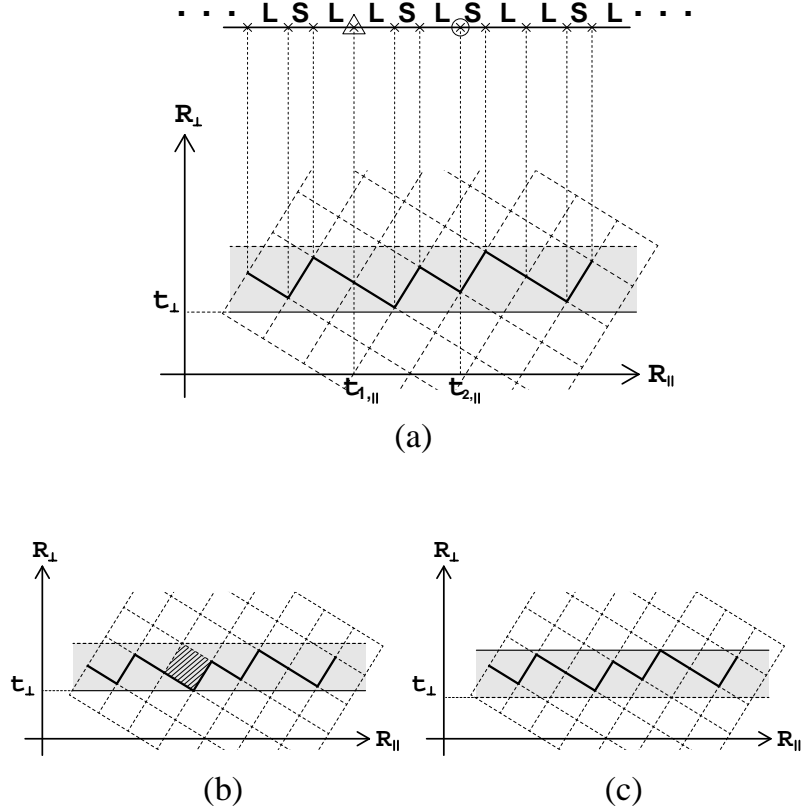,width=122mm}}
\caption{\small
The lifting of a Fibonacci chain (F-chain) into a 2D hyper-space.
For a given F-chain, we can construct a 1D tiling consists of two
prototiles $L$ and $S$ which can be lifted into a 2D square lattice
whose $x$-axis has the slope $-1/\tau$ with respect to the 1D tiling
on $\Rpar$. The embedded step (solid thick line) in the 2D lattice can
be covered by a strip parallel to
$\Rpar$ with width $\Delta = \cos\theta + \sin\theta$
when the position of the strip is well chosen.
The perp coordinate $t_{\perp}$ is given by the $\Rper$ coordinate of
the strip bottom. This value is uniquely determined for a given
infinite F-chain. In the upper part of (a), only a finite part of
the F-chain is shown. Therefore, the position of the strip which
covers the finite embedded step is not uniquely determined.
All three strips in (a), (b) and (c) cover the shown finite part but
correspond to different (infinite) F-chains.
The highest and the lowest strips, shown in (b) and (c)
respectively, corresponds to the singular F-chains.
The parallel coordinate $t_{\parallel}$ of the F-chain
depends on the choice of the vertex in the tiling.
}
\label{f-embed}
\end{figure}

An F-chain can be represented as a point in a 2-torus
${\mathbb T}^2$~\cite{Baake97}. Here we will show that
all F-chains in an equivalent class can be
represented as a leaf of the foliation on the torus.
First we construct a Fibonacci lattice (F-lattice) from
an F-chain.  An F-lattice is a 1D tiling
consists of two prototiles $L$ and $S$ whose arrangements
form an F-chain.  The ratio of the lengths of the two tiles,
$m(L)$ and $m(S)$ is given by $\frac{m(L)}{m(S)} = \tau$.
Figure~\ref{f-embed} shows an F-lattice (upper part of~(a))
and a way of lifting it into a 2D hyper-space which is a direct
product of two 1D spaces;  the ``parallel space'' $\Rpar$ and the
``perp-space'' $\Rper$.  The parallel space $\Rpar$ is a straight
line parallel to the F-lattice. The perp-space~$\Rper$
is the 1D space perpendicular to~$\Rpar$.

The coordinate of a vertex (F-lattice point; the boundary
between two given  consequent tiles) relative to any reference vertex
can be expressed in the form $n_{L}\, m(L) + n_S\, m(S)$,
where ($n_L$, $n_S$) $\in {\mathbb Z}^2$,
$m(S) = \sin\theta$ and $m(L)= \cos\theta$
with $\theta = \arctan(1/\tau)$.
Therefore, the vertices can be lifted into a square lattice of
$\mb{Z}^2$ as shown in the lower part of Fig.~\ref{f-embed}(a).
All pairs of adjacent vertices in the 1D tiling separated by
a tile $L$ or $S$ are mapped onto neighboring
vertices of a 2D square lattice in the $x$ or the $y$ directions
respectively where the $x$-axis has the slope $-1/\tau$.
The embedded step (thick solid line) in the 2D lattice by this lifting
can be covered by a strip parallel to the
$\Rpar$ with width $\Delta = \cos\theta + \sin\theta$
if the position of the strip is well chosen.

For a given (infinite) F-chain, the ``perpendicular space'' $\Rper$
coordinate $t_{\perp}$ of the strip (defined as the $\Rper$-coordinate
of the bottom boundary of the strip), which covers the embedded step
completely, is uniquely determined.\footnote{
In the upper part of Fig.~\ref{f-embed}(a), only a finite part of
the F-chain is shown. Therefore, the position of the strip which
covers the finite embedded step is not uniquely determined.
All three strips in the figure cover the finite embedded step.
The highest strip shown in (b), whose lower boundary passes the lattice point,
corresponds to the infinite F-chain whose index sequence
is $(101010\ldots)$ while the lowest strip in (c)
corresponds to $(01010\ldots)$.
Details will be discussed later.
}
Therefore, we can assign a $t_{\perp}$ value for a given F-chain.
The $\Rpar$ coordinate $t_{\parallel}$ of the chain
is not uniquely determined but depends on the choice of the vertex in
the 1D tiling. Figure~\ref{f-embed}(a) shows two different
values; $t_{1,\parallel}$ for the triangle vertex
and $t_{2,\parallel}$ for the circle vertex.
In section~\ref{Fibo_chain}, we mentioned that two sequences in $Z$
which are  equivalent by $\cR$ of Eq.~(\ref{e-equivR})
can be constructed from two different segments in the same F-chain. In
other words, two F-chains in an equivalence class can be considered as
a finite translation of each other. Since the translation in the
$\Rpar$ direction corresponds to the movement along the leaf in the
torus representation, all F-chains in an equivalent class
can be represented as the points on the same leaf on
the torus no matter what vertices we choose for $t_{\parallel}$.

Conversely, an F-lattice (hence an F-chain) can be obtained from a
2D square lattice by the projection methods.
The lattice sites of the 2D square structure can be projected onto
the 1D parallel space, $\Rpar$ at the slope $\tan \theta = 1/\tau$
with respect to the horizontal rows of the square lattices.
Since the slope of the line is irrational, the projection of all 2D
lattice points to $\Rpar$ form a dense set of points. If we restrict
projections on $\Rpar$ to the points confined within a strip
which is parallel to $\Rpar$ and has a cross section $\Delta$
in $\Rper$ equals to the perp-space projection of a square unit cell
($\Delta = \cos\theta + \sin\theta$),
then the projection to $\Rpar$ gives an F-lattice~\cite{Janot92B}.
The movement of the strip along the perp-space
$\Rper$ gives rise to rearrangement of tiles from one perfect
F-lattice to another and the strip is called the ``window'' of
the corresponding F-lattice.
In general, the windows should include one and only one boundary
to produce a perfect F-lattice.
Figure~\ref{f-embed} can be also used to illustrate the
``projection methods''. Now we first choose a window and select
the 2D lattice points which are in the window. Then the projection
of those lattice points into the $\Rpar$ space gives the vertices
of an F-lattice. The boundaries are irrelevant to the projected
structure unless they pass a 2D lattice point since all
vertices of the F-lattice are produced from the 2D lattice
points inside the window (Fig.~\ref{f-embed}(a)).
When a boundary of the window intersects with a 2D lattice point, so
does the other boundary as shown in Fig.~\ref{f-embed}(b) since the
width of the window equals to the perp-space projection of a square
unit cell. If the window included both boundaries, the projection
would produce an extra vertex from the 2D unit cell denoted by hatching.
On the other hands, if it excluded both, the projected lattice would
miss a vertex.
Therefore, the ``proper'' windows must include one and only one boundary.
If we include the lower boundary as shown in Fig.~\ref{f-embed}(b),
then we get the tile arrangement `$LS$' from the hatched unit cell,
while we get the `$SL$' arrangement for the other case. In other
words, there are two different F-lattices (and hence two different
F-chains) corresponding two different proper windows in spite of the
$\Rper$ position of the window is the same.

\begin{figure}[!t]
\centerline{\psfig{file=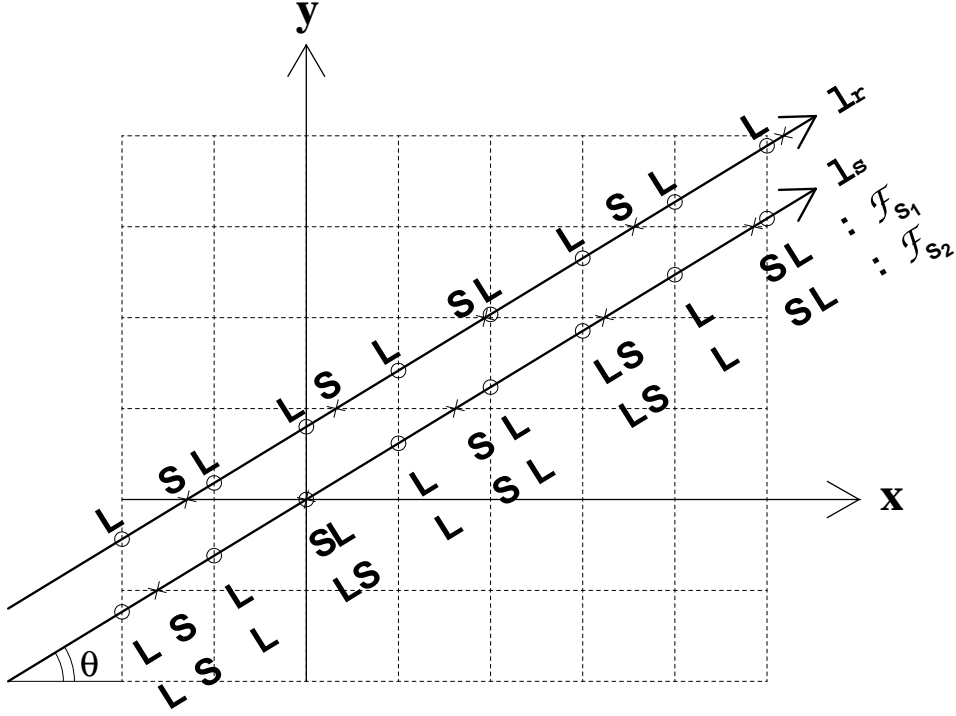,width=140mm}}
\caption{\small
An F-chain can be obtained from the
sequence of intersections between a line with
slop $1/\tau$ and the $x$ and $y$ axises.
A regular line $l_r$ intersects
$\cdots,y,x,y,y,x,y,x,y,y,x,y, \cdots$ axises
and corresponds to a unique F-chain,
$\cdots LSLLSLSLLSLL \cdots$.
The singular line $l_s$,
which passes through the origin, corresponds
to the two different F-chains, ${\cal F}_{s_1}$ and
${\cal F}_{s_2}$.
}
\label{f-proj}
\end{figure}

This ``singularity'' for the windows whose boundaries pass the lattice
points can be more clearly understood in the cut-procedure.
A leaf on the torus can produce an F-chains (but not an F-lattice)
directly (instead of going through a strip or a window) in the
cut-procedure. Figure~\ref{f-proj} illustrates a way to get an F-chain
by this method from a square lattice in a 2D ``hyper-space''.
We consider a 2D square lattice and the lines with the slop of
$1/\tau$. (A straight line in a square lattice can be
considered as a representation of a leaf on the torus in
the ``extended'' scheme.) We can produce an F-chain associated to the
line in the following way. If the line intersects the $y$-axis, we give the
segment ``$L$'' while we assign the segment ``$S$'' when the line
intersects with the $x$-axis.  For example, the line $l_r$ in
Fig.~\ref{f-proj} intersects $\cdots,y,x,y,y,x,y,x,y,y,x,y, \cdots$
axises and hence
produces an F-chain ${\cal F}_r= \cdots LSLLSLSLLSLL \cdots$.
The correspondence between a straight line and an F-chain
is one-to-one except the ``singular'' line $l_s$ which passes
the origin. For the singular case, the three lines, the parallel
line $l_s$, the $x$-axis and the $y$-axis meet at a point
(at the origin). Since $l_s$ meets both the $x$ and the $y$ axises at
the same time, a pair of segments (one $S$ and one $L$ segments)
should be assigned at the origin. However, assigning two
segments at the same point is impossible.
This ``singularity'' can be resolved by moving the line $l_s$
infinitesimally. If we move $l_s$ slightly upward,
it first meets the $x$-axis and then the $y$-axis
and `$SL$' is assigned at the origin. Therefore,
the parallel line $l_s$ produces the F-chain ${\cal F}_{s_1}$ in
Fig.~\ref{f-proj} when it is moved upward infinitesimally.
In contrast, if we move $l_s$ slightly downward,
it meets the $x$-axis first and then meets the $y$-axis.
Therefore, `$LS$' is assigned at the origin and we have
${\cal F}_{s_2}$ in this case.

Now, the space of F-chains can be parameterized by the y-axis in
$\mathbb R$ since the line $y=\frac{1}{\tau} x+b$ is determined
by the $y$-intercepts. As in the leaf space of the Kronecker
foliation, two F-chains that correspond to two lines
$\frac{1}{\tau}x+b$ and $\frac{1}{\tau}x+b'$ are in the same
equivalence class if $b-b'=n/\tau$. This is because the
arrangements of intersections from the two lines,
hence the two corresponding F-chains, are the same up to the finite
translation (by $n\sqrt{\tau^2+1}/\tau$) when
$b-b'=n/\tau$.
If two F-chains ${\cal F}$ and ${\cal F}'$ are the same up to
a finite translation, they are in the same class
by the equivalence relation $\cR$ of Eq.~(\ref{e-equivR}).
Let $(a_n)$ and $(a'_n)$ be index sequences of the two
F-chains with $(a_n) = i({\cal F},\alpha)$
and $(a'_n) = i({\cal F'},\alpha')$
where $\alpha$ and $\alpha'$ are the segments at the origins of
${\cal F}$ and ${\cal F'}$
respectively. Then, there is a segment
$\alpha'' \in {\cal F}$ within a finite distance
from the origin such that
$(a''_n) = i({\cal F},\alpha'')$
be the identically same as $(a'_n)$
since ${\cal F'}$ is a finite translation of ${\cal F}$.
Now, we have an integer $M$ such that $a_n = a'_n$ for all $n>M$
since the inflation will make two segments $\alpha$ and
$\alpha''$ belong to the same segment in ${\cal F}_n$ for sufficiently
large $n$.

We have shown that the space of F-chains is the same
as the leaf space of Kronecker foliation except that
the singular leaf which corresponds to the two F-chains.
One may think that the two F-chains corresponding to the
singular leaf are the same class. The only difference
between two chains are at the origin which can be removed
by a local surgery; we can obtain one chain from the
other by flipping  one pair of segment at the origin.
Furthermore, one is a mirror image of the other and related by
a 180 degree rotation. However, they are not in the same class by the
equivalence relation $\cR$ of Eq.~(\ref{e-equivR}).
If we construct the index sequences of the two F-chains from the
segment at the origin, one chain corresponds to the index sequence
$(a_n)=(0101010101\ldots)$
and the other corresponds to
$(a'_n)=(1010101010\ldots)$ (see section~4).
In other words,
$ a_n = \delta_{n,2k} $
for one chain while
$ a'_n = \delta_{n,2k+1} $
for the other chain. Clearly $a_n$ and $a'_n$ are different for all $n$
and they are in different classes.

Since we have two different F-chains on the singular leaf (which is
only one leaf on the Kronecker foliation) on $\mathbb T^2$, we have a
surjective map from the space of F-chains to the space of leaves.
Both spaces have trivial topology and the map is open and continuous.
Now, the surjectiveness corresponds to that
the map from the C*-algebra
of leaves of foliation (noncommutative torus) to the C*-algebra of
F-chains (we already showed in section 2 that it is an AF-algebra) is
injective. In this sense our discussion above can be seen as a
dual picture of the embedding of  noncommutative torus into a
certain type of AF-algebra.

In the following section, we construct such an AF-algebra by
introducing an extended space of leaves which is isomorphic to
the space of F-chains.


\section{An extended space of leaves}

Since the map from the space of F-chains to the leaf space of the
Kronecker foliation is surjective, we cannot retrieve all F-chains
from the leaves on ${\mathbb T}^2$. However, there is only one leaf
which corresponds to more than one class of F-chains. Furthermore,
this singular leaf corresponds to only two classes of
F-chains. Therefore, if we assign one class of F-chain to every leaf
(including the singular leaf), all F-chains except only one class of
F-chains are obtained from the leaf space. For example, if
we assign ${\cal F}_{s_2}$ in Fig.~\ref{f-proj} to the singular leaf,
then ${\cal F}_{s_1}$-class is ``missing'' but all other F-chains
are in the leaf space on $\mathbb T^2$. In this section, we show that
an extended leaf space, which is isomorphic to the space of F-chains,
is naturally obtained if we construct the equivalence relations on the
leaf space using the equivalence relation~$\cR_n$ of the finite
subsequences of the index sequences given by Eq.~(\ref{e-equivRn}).
In the limit of the length of the subsequences goes to infinity, we
get the extended leaf space which is the sum of two spaces; the leaf
space on~$\mathbb T^2$ and the space consists of one leaf
corresponding the ``missing'' class.

\begin{figure}[!t]
\centerline{\psfig{file=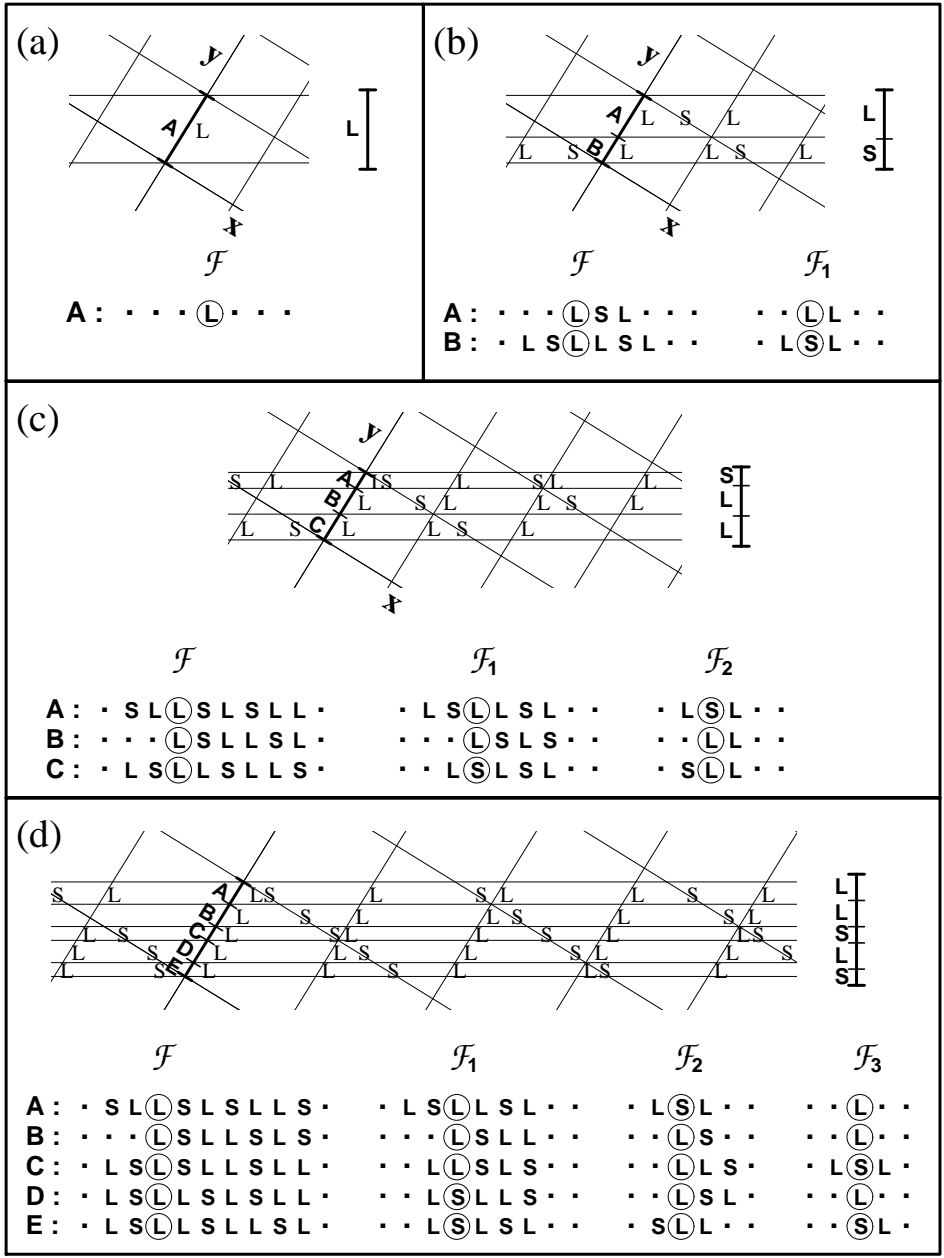,width=140mm}}
\caption{\small
A sequence of partitions of the transversal
(1,0) using the cut-procedure and the inflations
of the F-chain.
$\cF_{n+1}$ is the inflation of $\cF_n$ as in
Fig.~\ref{f-inf}.
}
\label{f-AF}
\end{figure}

In section~\ref{Fibo_chain}, the index subsequences of the
F-chains are constructed using the inflation and the AF-algebra is
introduced as an inductive limit of the finite algebra on the space of
the finite subsequences of the index sequences. There were only two
sets of equivalence classes on the space of the finite subsequences
since the last entries of sequences $a_n$ in $Z_n$ can have only two
values, either 0 or 1. Here, we consider the space of straight lines
in the cut-procedure of Fig.~\ref{f-proj}.
Except for the singular lines which passes the lattice points,
each line produces one and only one F-chain. The equivalence relation
between the lines is constructed according to the equivalence
relation~$\cR_n$ for their F-chains (Eq.~(\ref{e-equivRn})).
As before, the straight lines are parameterized by the $y$-intercepts.
To define the equivalence relation between the lines,
we consider the index sequences of their F-chains
$i({\cal F}(b),\alpha)$ where $\alpha$ is the segment at
$x=0$ (at the $y$-axis) and ${\cal F}(b)$ is the F-chain
corresponds to the line $y=\frac{1}{\tau} x+b$.
The index $a_0$ of the original (uninflated) F-chain
${\cal F}$ is $0$ for every $b\in W_0 := (0, 1)$
since the type of the segment at the origin is always $L$ by the
definition of the cut-procedure
(at $x=0$, the leaf $y=\frac{1}{\tau}x+b$ always cut the $y$-axis).
Figure~\ref{f-AF}(a) illustrates this;
any $b \in W_0$ corresponds to the $L$ segment in ${\cal F}$ at
the origin. However, the type of the segments at the origin
in the inflated F-chains ${\cal F}_1$ can be both $L$ and
$S$ depending on the value of $b$.
As shown in Fig.~\ref{f-AF}(b), the segment arrangement
in $\cF$ at the origin is $LS$ for
$b\in A = (1/\tau^2,1)$,
and $LL$ for $b\in B = (0,1/\tau^2)$.
Since $LS$ becomes $L$ and $L$ becomes $S$ by inflation,
the segment type in $\cF_1$ at the origin is $L$ for $b\in A$
and $S$ for $b\in B$.
Therefore, $W_0$ is partitioned by two open intervals, $A$ and $B$
and a boundary point $b_1=1/\tau^2$ for $n=1$.
Let us denote the union of the two open intervals by $W_1$;
$W_1 = (0,1/\tau^2)\cup(1/\tau^2,1) = W_0 - b_1$.
Note that the boundary point $b_1$ is given by the intersection
between $W_0$ and the line which passes the 2D lattice
point (1,1). For this boundary line $y=\frac{1}{\tau}x+(1-1/\tau)$,
the segment type at the origin of $\cF_1$ is not well defined,
that is, $b_1$ is the singular point for $\cF_1$.
Similarly, we can partition $W_1$ by considering
the doubly inflated F-chains ${\cal F}_2$. Since an $S$-segment
in ${\cal F}_1$ becomes an $L$-segment in ${\cal F}_2$,
the interval $B$ in Fig.~\ref{f-AF}(b) is not
divided for $n=2$ (interval $C$ in Fig.~\ref{f-AF}(c)).
The interval $A$ in Fig.~\ref{f-AF}(b) which represents
the class of $a_1 = 0$ is divided by two intervals by the
line which passes the 2D lattice point (2,2). Therefore, we get
the three intervals,
$W_2 = (0,1/\tau^2)\cup(1/\tau^2,2/\tau^2)\cup(2/\tau^2,1)$
for $n=2$ as shown in Fig.~\ref{f-AF}(c).
In general, an interval corresponding to $a_n=1$ becomes an interval
corresponding to $a_{n+1} = 0$ in the next step while
an interval corresponding to $a_n=0$ will be divided as
two neighboring intervals, one for $a_{n+1} = 0$ and the other
for $a_{n+1} = 1$.
Therefore, the partitioned interval for the $n$th inflated chain,
which will be denoted by $W_n$, is given by
the union of $f_{n+2}$ intervals with $f_{n+1}$ $L$-intervals and
$f_{n}$ $S$-intervals;
\begin{eqnarray}
 W_n &=& W_{n,L} + W_{n,S}, \nonumber
\end{eqnarray}
where
\begin{eqnarray}
 W_{n,L} = \sum_{k=1}^{f_{n+1}}  I_{n,L_k},
 &\ \ &
 W_{n,S} = \sum_{k=1}^{f_{n}}  I_{n,S_k}. \nonumber
\end{eqnarray}
Here, $I_{n,L_k}$ ($I_{n,S_k}$ resp.) is the $k$th interval of
$L$-type ($S$-type resp.) in $W_n$ and
$f_k$ is the Fibonacci number introduced in section~2.
The lengths of the $L$ and the $S$-intervals in $W_n$ are
$1/\tau^n$  and $1/\tau^{n+1}$ respectively.

\begin{figure}[!t]
\centerline{\psfig{file=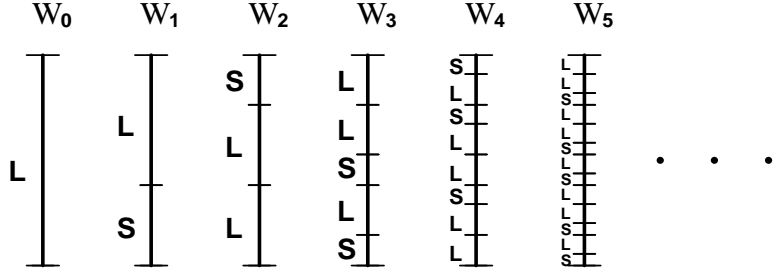,width=120mm}}
\caption{\small
A sequence of partitions of the transversal for the
construction of a sequence of finite dimensional algebras.
An $L$-interval in $W_n$ is divided by an $L$- and
an $S$-interval in $W_{n\!+\!1}$ but an $S$-interval
becomes an $L$-interval without partition in the next step.
For even $n$, the lower part of an $L$-interval becomes
an $S$-interval in the ($n\!+\!1$)th inflation while the upper part
becomes an $S$-interval for odd $n$.
}
\label{f-AFW}
\end{figure}

Figure~\ref{f-AFW} shows the arrangement of
$I_{n,L_k}$ and  $I_{n,S_k}$ in $W_n$. From Fig.~\ref{f-AF}, we see
that the intervals are divided by the
lines which pass the 2D lattice point $(r,s)$
such that $0<s - r/\tau < 1$.
Let us arrange such lattice points according to the ``parallel''
distance
$d(r,s) = \frac{\tau}{\tau+2} (r + s/\tau)$
and denote them as $P_k = (r_k,s_k)$ where
$ d(r_{k'},s_{k'}) < d(r_k,s_k)$ for $k'<k$.
Now, let $l_k$ be the line which passes the lattice
point $P_k$. Then the $f_{n+1}$ lines,
$l_{f_{n+2}}, \cdots, l_{f_{n+3}-1}$,
divide the $f_{n+1}$ $L$-intervals in $W_n$.
For example,
$l_1$ which passes $P_1=(1,1)$ divides the $L$-interval in $W_0$,
$l_2$ which passes $P_2=(2,2)$ divides the $L$-interval in $W_1$,
and
$l_3$ and $l_4$ which pass $P_3=(3,2)$ and $P_4 = (4,3)$
divide the $L$-intervals in $W_2$.
Note that $l_1$ divides the $L$-interval of $W_0$
such that the lower part of it becomes an $S$-interval in $W_1$
while $l_2$ divides the $L$-interval in $W_1$
such that the upper part of it becomes an $S$-interval in $W_2$.
The $L$-intervals in $W_2$ are divided by the lines $l_3$
and $l_4$ as the way that the $L$-interval of $W_0$ was divided.
In fact, the lower part of an $L$-interval becomes an $S$-interval for
even $n$ while the upper part of an $L$-interval becomes
an $S$-interval for odd $n$ by an
inflation.\footnote{
This can be shown by two steps;
(1) All $L$-intervals in $W_n$ are divided in the
same pattern.
(2) A particular $L$-interval in $W_n$ is divided as the way
mentioned above.
To prove~(1), let $l_k$ and $l_{k'}$ be the two boundary lines
of an $L$-interval in $W_n$ and pass
the lattice points $P_k = (r_k,s_k)$
and $P_{k'} = (r_{k'},s_{k'})$ respectively.
The irrationality of the slop guarantees the same
$|\Delta r| = |r_k - r_{k'}|$ and
$|\Delta s| = |s_k - s_{k'}|$
for all $L$-intervals in $W_n$;
all of them are given by
$|\Delta r| = f_n$ and
$|\Delta s| = f_{n-1}$
due to the relations
$f_{n-1} - f_n/\tau = (-1)^n\frac{1}{\tau^n}$.
Now, the $L$-intervals in $W_n$ can be mapped to pairs of
2D lattice points which are identical up to a lattice
translation. In other words, relative arrangements of the lattice
points from the ``boundary'' lattice points
are the same for all $L$-intervals in $W_n$ and
hence divided in the same pattern by the boundary
lines in the following inflations.
The statement~(2) can be shown with the bottommost
interval for even $n$ and the topmost interval for odd~$n$
with an inductive way. Here we outline the proof for
even $n$. The odd $n$ case can be proven similarly.
For $n=2$, the bottommost interval is an $L$-interval
and the upper boundary is given by the line
which passes the lattice point $(1,1)$.
Let the bottommost interval of $n = 2k$ case is
an $L$-interval. Then the size of the interval is
$\tau^{-n}$ and the upper boundary line passes
the lattice point $(f_n,f_{n-1})$. This interval will be
divided by the lines which pass the lattice points
$(r,s)$ such that $ 0 < r - s/\tau < \tau^{-n}$.
Since $f_n/f_{n-1}$ is the ``best'' rational approximant
of $\tau$~\cite{Olds63B}, the lattice point
$(f_{n+2},f_{n+1})$ has the smallest parallel
distance and hence the line which pass it
divides the interval first. The ``perpendicular''
distance of this lattice point is
$f_{n+1}-f_{n+2}/\tau = \tau^{-(n+2)}$
and we see that the lower partition of the interval
becomes an $S$-interval in $W_{n+1}$.
Furthermore, this implies the bottommost interval
for $n+2$ is the $L$-type and the above argument
can be applied inductively for all even~$n$.
}
Successive application of these processes produces
the sequence of partitions shown in
Fig.~\ref{f-AFW} and the two groups of intervals
$W_{n,L}$ and $W_{n,S}$ can be obtained for every~$n$.

~From the construction above, we can see that the set of all intervals
in $W_n$ is isomorphic to $Z_n$ in section~2.
For a given sequence $z_n=(a_1,a_2,\ldots,a_n)$
in $Z_n$, we can choose an interval in $W_n$ in
the following manner.
First, choose the interval $L$ ($S$ resp.) in $W_1$ if $a_1 = 0$
($a_1=1$ resp.). Then choose the interval $L$ ($S$ resp.)
in $W_2$, which is a subinterval of the previously chosen one
if $a_2 = 0$ ($a_2=1$ resp.). For $a_n = 0$ ($a_n=1$ resp.),
choose the interval $L$ ($S$ resp.) which is a
subinterval of the chosen interval in $W_{n-1}$.
Then, there is always an interval in $W_n$ for a given
sequence in $Z_n$ since an interval $S$ in
$W_k$ becomes an interval $L$ in $W_{k+1}$.
Conversely, an interval in $W_n$ can be indexed by a sequence
in $Z_n$ by recording the types of the intervals
in $W_k$ (for $k=1,\cdots,n$) which the chosen interval belongs to.
Now we can identify
$W:=\lim_{n\rightarrow\infty} W_n$ with the set
$Z = \lim_{n\rightarrow\infty} Z_n$ and hence
$W$ is  the set of all
F-chains. Note that  the boundary points excluded from the $n$th
partitioned interval, $W_0 - W_n$, are the first $f_{n+1}$ orbit
points of the irrational rotation $-1/\tau$ from
$1-1/\tau = 1/\tau^2$. In other words,
the lines through lattice
points corresponds to the orbit of the rotation defined by the
diffeomorphism $\psi_\tau$ introduced in section 3. In fact
the construction of $W$ is exactly the same
as that of the exceptional minimal set for the suspension of
diffeomorphism $\psi_\tau$. Thus the set $W$ or the set of all
F-chains is the exceptional minimal set and also the set $Z$ is
homeomorphic to the Cantor set as asserted
in~\cite{Connes94B,bigatti98}.

Now, let us give an equivalence relation $\tilde{\cR}_n$ on $W_n$ as
$\cR_n$ of Eq.~(\ref{e-equivRn}) on $Z_n$. Then the set of equivalence
classes, $\tilde{X}_n = W_n/\tilde{\cR}_n$ has only two elements,
$W_{n,L}$ and $W_{n,S}$ which have $f_{n+1}$ and $f_{n}$ intervals
respectively. By taking these $f_{n+1}$ and $f_{n}$ intervals as the
bases of the $W_{n,L}$ and $W_{n,S}$ classes respectively, we recover
the sequence of finite algebras described in section~\ref{Fibo_chain}.
In the limit of $n$ goes to infinity, we obtain an AF-algebra which is
the same as in section~2.

We now show that the space
$\tilde{X} = \lim_{n\rightarrow\infty} \tilde{X}_n$,
which is isomorphic to the space of F-chains $X$,
is given by the quotient space obtained from $W$ mod out by the
``leaf equivalence relation''; $b\sim b'$ iff $b-b'=n/\tau$ for
some integer $n$, and call $\tilde{X}$ as ``extended leaf space''.
An important consequence of the partition sequence
of Fig.~\ref{f-AFW} is that all $L$-intervals in $W_n$ are
divided in the same pattern in $W_{n+1}$ (see footnote~(1) also).
Furthermore, all $S$-intervals in $W_n$ become the $L$-intervals in
$W_{n+1}$. Therefore, all intervals of the same type in $W_n$
are divided in the same pattern in $W_m$ for all $m>n$.
This observation provides the relation between two
points, $b_z$ and $b_{z'}$, in an equivalence class
which can be indexed by two sequences $z = (a_k)$
and $z'=(a'_k)$ with $a_m = a'_m$ for all $m \ge n$.
If $I_n$ ($I'_n$ resp.) is the interval in $W_n$,
which $b_z$ ($b_z'$ resp.) belongs to, the relative
distance from a reference point (say, the center) of
$I_n$ to $b_z$ is the same as that from the center of
$I_n'$ to $b_{z'}$ because $a_m = a'_m$ for all $m \ge n$.
Since the lengths of $L$-intervals and $S$-intervals
in $W_n$ are $\tau^{-n}$ and $\tau^{-(n+1)}$ respectively,
the distance between $b_z$ and $b_{z'}$ is given by
\begin{eqnarray}
 b_z - b_{z'}
   &=&  k_L \tau^{-n}
      + k_S \tau^{-(n+1)} \nonumber \\
   &=& (-1)^n \Big[
         (k_S f_{n+1} - k_L f_{n}) \frac{1}{\tau}
       + (k_L f_{n-1} - k_S f_{n})
       \Big]  \nonumber \\
   &=& \frac{k_1}{\tau} - k_2
\label{e-equiP}
\end{eqnarray}
with integers
$k_1 = (-1)^n (k_S f_{n+1} - k_L f_n)$
and
$k_2 = (-1)^n (k_L f_{n-1} - k_S f_{n})$.
Here $k_L$ and $k_S$ are the number of $L$ and $S$-intervals
between the two chosen intervals in $W_n$ and we used the relation
$\tau^{-k} = (-1)^k (f_{k-1} - f_k/\tau)$.
These are the exactly the same condition for the
same leaf on the torus in section~\ref{s-TR}.

We should note that the space obtained by the limit of the above
procedure is not the space of leaves of the Kronecker foliation on
$\mathbb T^2$. If we follow the very bottom intervals of $W$ in
Fig.~\ref{f-AFW}, we get the F-chain whose index sequence is given by
($010101010\ldots$) while we get ($001010101\ldots$) when we follow
the very top intervals. Therefore, the limit points of these
two sequences, 0 and 1 represent different classes.
In the foliation on the torus, above two leaves had to be
identified since 0 and 1 are identical point in $S_1$.
This was the reason that the singular leaf on the torus
corresponds two distinct F-chains.


\section{Concluding remarks}

In this paper, we studied the space of F-chains from the perspective
of noncommutative geometry. We defined the equivalence relation of the
F-chains based on their index sequences and the space of F-chains $X$
is given as the quotient set $Z/\cR$ of all F-chains $Z$ divided by the
equivalence relation $\cR$ of Eq.~(\ref{e-equivR}). This space is
exactly the same as the space of Penrose tiling considered by
Connes~\cite{Connes94B}. From the calculation of its
$K$-theory, we know that the
$K_0$-group of Penrose tiling (and hence F-chains) is isomorphic to that
of the noncommutative torus. Furthermore, an F-chain can be
parameterized as a point on the torus
${\mathbb T}^2$~\cite{Baake97}. These facts suggest a strong connection
between the noncommutative torus and the space of F-chains.  However,
a $C^*$-algebra on the space of F-chains cannot be a noncommutative
torus. From the Connes' work~\cite{Connes94B}, we know that the
construction of a nontrivial algebra on the space of F-chains gives
rise to an AF-algebra whose $K_1$ vanishes while the $K_1$ of the
noncommutative torus does not.

Here, we studied the exact relationship between the noncommutative
torus and the AF-algebra on the space of F-chains. Using the torus
representation and the cut-procedure, we found a surjective map from
the space of F-chains to the space of leaves on Kronecker foliation.
The surjectiveness of the map and the embedding of the $C^*$-algebra
of noncommutative torus to an AF-algebra was explicitly shown by
considering a sequence of finite-algebra constructed on the finite
partitions $I_{n,L_k}$ and $I_{n,S_k}$, the quotient space
$W_n/\cR_{n}$.
In the limit of $n$ goes to infinity, the quotient space is
identified with
the space of $F$-chains. This is  a dual approach to
the way of
embedding of the $C^*$-algebra of the space of leaves on Kronecker
Foliation into the $C^*$-algebra of the space of F-chains.

We think the current method of finding the relationship between the
space of F-chains and the space of leaves of Kronecker Foliation can
be applied to the space of Penrose tilings.
As an F-chain can be represented as a leaf on ${\mathbb T}^2$,
which is a line parallel to the 1D space spanned by a 2D vector
$(\cos\theta, \sin\theta)$ with $\tan\theta = 1/\tau$, a Penrose
tiling can be represented as a plane on
${\mathbb T}^5$
which are parallel to the 2D space spanned by two 5D vectors,
$(1, c_1, c_2, c_3, c_4)$ and $(1, s_1, s_2, s_3, s_4)$ where
$c_k = \cos\left(\frac{2\pi}{5}\,k\right)$
and
$s_k = \sin\left(\frac{2\pi}{5}\,k\right)$~\cite{Janot92B}.
Recall that a leaf on
${\mathbb T}^2$, $y=\frac{1}{\tau}\,x + b$
can be parameterized by the $y$-cut $b$ by identifying the position of
a leaf as the point of $x=0$ on the leaf.
Similarly, a Penrose tiling can be parameterized by
the position of the origin of the plane.  By introducing the
equivalence relation between the positions of the planes
according to the equivalence relationship of their Penrose tiling,
we can construct a space of wrapping 2D planes (``2D leaf'') in the 5D
space. From the identity between the space of F-chains and that of
Penrose tiling, one may expect that this space should be
very similar to the space of leaves of Kronecker foliation.
However, a preliminary study shows that this may not be the case.
The properties of the singular plane in this space, which passes
through the origin of the 5D space, may behave quite differently from
the singular leaf of F-chains. The singular plane corresponds to 5
different Penrose tilings but their index sequences are the same and
given by $(0,0,0,\ldots)$. Therefore, all of them are in the same
equivalence class unlike the F-chains from the singular leaf.
Further work on this issue may have practical application
for the study of quasicrystal structures. The decapod defects, which
can be a seed for rapid quasicrystal growth~\cite{Onoda88}, are known
to be related to the singularity of the plane which passes the origin of
5D space~\cite{Bruijn81b}. If future studies establish the role of the
higher dimensional singular ``leaf'' in the hyper lattice space for
the space of general quasiperiodic tilings, they may provide a new
clue to solve the old puzzle of the topological character of the
decapod defects~\cite{Janot92B}.

We hope the current study on the space of F-chains provides a motive
for the proliferation of noncommutative geometrical approaches
for the properties of Penrose tiling and quasicrystals.
At the moment, the progress on the dynamical properties of the
quasicrystalline structure seems to be slow.
There have been a great deal of studies on the dynamical properties on
the 1D quasiperiodic systems but a very little innocuous extensions to
the higher dimensional quasicrystalline structure have been
successfully made. We speculate that the study on the dynamics
of 1D quasiperiodic system from the noncommutative geometrical
aspect may provide a new tool for the investigation to the
dynamics of quasicrystals.  As shown in this paper, both
the space of F-chains (1D quasiperiodic systems) and the
space of Penrose tiling (2D quasicrystalline lattice) show
the same non trivial structure only with a noncommutative
geometrical approach.

In summary, we show that the noncommutative torus can be obtained from
the space of Fibonacci chains via foliation. We hope that this
understanding will help to enhance the understanding of the dynamics
of the quasicrystalline structure in future.

\vspace{5mm}
\begin{center}
{\large \bf Acknowledgments} \\
\end{center}
\vspace{1mm}
This work was supported by
Korea Research Foundation,
Interdisciplinary Research
Project 1998-D00001.
Eunsang Kim was also supported
by BK 21.
\\


\begin{thebibliography}{10}
\bibitem{expl_cantor}
$Z$ is homeomorphic to the Cantor set. For details, see Ref.~\cite{bigatti98}.
\end{thebibliography}
\end{document}